\documentclass[aps,prl,twocolumn,superscriptaddress,showpacs]{revtex4}

\begin{document}

\title{No-Signalling Bound on Quantum State Discrimination}

\author{Sarah Croke}
\email{sarah@phys.strath.ac.uk}
\affiliation{Department of Physics, SUPA, University of Strathclyde, Glasgow G4 0NG, UK}
\affiliation{Department of Mathematics, University of Glasgow, Glasgow G12 8QW, UK}
\author{Erika Andersson}
\author{Stephen M. Barnett}
\affiliation{Department of Physics, SUPA, University of Strathclyde, Glasgow G4 0NG, UK}

\pacs{03.65.Ud, 03.67.Hk, 03.65.Ta}

\date{\today}

\newcommand{\bra}[1]{\langle #1|}
\newcommand{\ket}[1]{|#1\rangle}
\newcommand{\braket}[2]{\langle #1|#2\rangle}

\begin{abstract}
Quantum correlations do not allow signalling, and any operation which may be performed on one system of an entangled pair cannot be detected by measurement of the other system alone.  This no-signalling condition limits allowed operations and, in the context of quantum communication, may be used to put bounds on quantum state discrimination.  We find that the natural figure of merit to consider is the confidence in identifying a state, which is optimised by the maximum confidence strategy.  We show that this strategy may be derived from the no-signalling condition, and demonstrate the relationship between maximum confidence measurements and entanglement concentration.
\end{abstract}

\maketitle

Although entanglement appears to allow particles which are separated in space to influence one another instantaneously \cite{asp}, it has been shown that this cannot be used to transmit information superluminally \cite{ghi}, thus reconciling quantum mechanics and special relativity.  Any operation which can be performed on a quantum system must therefore be compatible with no-signalling, and imposing this condition allows us to put limits on certain tasks, for example those obtained on the fidelity of quantum cloning machines \cite{clone}, and on error-free discrimination between non-orthogonal quantum states \cite{amb}.  As a consequence, it has been suggested that no-signalling should be given the status of a physical principle, used to restrict quantum mechanics and possible extensions of it \cite{pr,mas}, and, interestingly, a quantum key distribution protocol has been devised which is provably secure against attacks from an eavesdropper limited only by no-signalling \cite{barr}. 

In state discrimination, in addition to the limits derived on error-free discrimination \cite{amb}, the no-signalling condition has been used to bound the minimum probability of error in discriminating between two non-orthogonal states \cite{hwang}.  In this paper we show that, in the problem of discriminating between a set of non-orthogonal states $\{ \ket{\psi_i} \}$, the no-signalling condition leads us to consider the conditional probability $P(\psi_j|\omega_j)$, where $\omega_j$ denotes the measurement outcome which leads us to choose state $\ket{\psi_j}$.  This is the figure of merit optimised by the recently introduced maximum confidence strategy \cite{maxconf}.  We will show that no-signalling gives a direct way to obtain a tight bound on this figure of merit, without reference to Probability Operator Measures (POMs) \cite{hel}.  Further, by considering when the bound is reached, we will show that it is then possible to derive the POM describing the optimal measurement using only no-signalling.  We illustrate this by means of an example, and discuss the relationship to entanglement concentration.

Suppose that two systems, referred to as left and right, are entangled in the state
\begin{equation}
\ket{\Psi} = \sum_{i=0}^{N-1} p_{i}^{1/2} \ket{\psi_{i}}_{L} \ket{i}_{R}
\label{psi}
\end{equation}
where the states $\{ \ket{\psi_i}_L \}$ of the left system lie in a $D \leq N$ dimensional space, and the set $\{ \ket{i}_{R} \}$ forms an orthonormal basis for the right system.  In a quantum communication protocol, the states $\{ \ket{\psi_i}_L \}$ may be prepared with \emph{a priori} probabilities $\{ p_i \}$ by performing a projective measurement onto the orthogonal states $\{ \ket{i}_R \}$ of the right system.  Further, suppose that a measurement is subsequently performed on the left system to discriminate between these states.  After measurement, the probability that the initial state was $\ket{\psi_{j}}_{L}$ may be interpreted as the probability that the measurement on the right system gave the outcome corresponding to state $\ket{j}_{R}$.  As operations on different systems commute, this interpretation is valid regardless of the order in which the measurements are performed.  Thus, if a measurement performed on the left system of the entangled state $\ket{\Psi}$ yields outcome $\omega_{j}$, the reduced density operator of the right system is transformed to $\hat{\rho}_{R|j}$ and the probability $P(\psi_{j}|\omega_{j})$, the confidence in identifying state $\ket{\psi_j}_L$, may be expressed:
\begin{equation}
P(\psi_{j}|\omega_{j}) = _{R}\bra{j} \hat{\rho}_{R|j} \ket{j}_{R}.
\label{confidence}
\end{equation}
It is therefore natural to consider conditional probabilities of this form.  The no-signalling condition restricts $\hat{\rho}_{R|j}$, and therefore may be used to put a bound on the above probability.  The Schmidt decomposition \cite{nie} of the entangled state $\ket{\Psi}$ may be written
\begin{equation}
\ket{\Psi} = \sum_{i=0}^{D-1} \lambda_{i}^{1/2} \ket{\lambda_{i}}_{L} \ket{\lambda_{i}}_{R}
\label{schmidt}
\end{equation}
where $\sum_{i} \lambda_{i} = 1$ and $_{L}\braket{\lambda_{i}}{\lambda_{j}}_{L} = _{R}\braket{\lambda_{i}}{\lambda_{j}}_{R} = \delta_{ij}$.  Thus the set $\{ \ket{\lambda_{i}}_{L} \}$ is an orthonormal basis of the left system, and we can construct $N-D$ states $\ket{\Psi_{i}^{\perp}}_{R}$ of the right system alone, such that $\{ \ket{\lambda_{i}}_{R}; i=0,...D-1, \ket{\Psi_{i}^{\perp}}_{R}; i=0...N-D-1 \}$ forms an orthonormal basis of the right system, and $_{R}\braket{\Psi_{i}^{\perp}}{\Psi} = 0$.  The no signalling condition implies that no operation performed on the left system may be detected by observation of the right system alone.  In particular, after the measurement described above is made on the left system, the state of the right system is described by the density operator $\hat{\rho}_{R|j}$.  Clearly if $_{R}\bra{\Psi_{i}^{\perp}} \hat{\rho}_{R|j} \ket{\Psi_{i}^{\perp}}_{R} \neq 0$ for any $\ket{\Psi_{i}^{\perp}}_{R}$, there is a finite probability that the operation could be detected by a measurement on the right system alone.  Such a transformation would allow the possibility of superluminal communication, and is therefore forbidden.  The $N$-dimensional right system is thus confined to a $D$-dimensional subspace, $\mathcal{H}_{RD}$, due to its entanglement with the left system.  The projector onto this subspace is
\begin{equation}
\hat{P}_{D} = \hat{I}_{R} - \sum_{i=0}^{N-D-1} \ket{\Psi_{i}^{\perp}}_{R R}\bra{\Psi_{i}^{\perp}} = \sum_{i=0}^{D-1} \ket{\lambda_{i}}_{R R}\bra{\lambda_{i}}.
\end{equation}
It is useful to write this in the form
\begin{equation}
\hat{P}_D = {\rm Tr}_L (\hat{\rho}_L^{-1/2} \ket{\Psi} \bra{\Psi} \hat{\rho}_L^{-1/2}),
\label{PD}
\end{equation}
where $\hat{\rho}_L$ is the reduced density operator of the left system,
\begin{equation}
\begin{array}{ccl}
\hat{\rho}_L = {\rm Tr}_R(\ket{\Psi} \bra{\Psi}) &=& \sum_i \lambda_i \ket{\lambda_i}_{L \, L} \bra{\lambda_i}, \\ &=& \sum_i p_i \ket{\psi_i}_{L \, L} \bra{\psi_i}.
\end{array}
\end{equation}

We wish to maximise ${}_R \bra{j} \hat{\rho}_{R|j} \ket{j}_R$ subject to the constraint that $\hat{\rho}_{R|j}$ lies in the subspace $\mathcal{H}_{RD}$.  Clearly then $\hat{\rho}_{R|j}$ is simply the projection of $\ket{j}_{R \, R} \bra{j}$ onto this subspace
\begin{equation}
\hat{\rho}_{R|j} = \frac{\hat{P}_D \ket{j}_{R \, R} \bra{j} \hat{P}_D}{{}_R \bra{j} \hat{P}_D \ket{j}_R},
\label{rhoRj}
\end{equation}
giving
\begin{equation}
[P(\psi_{j}|\omega_{j})]_{max} = _{R}\bra{j} \hat{P}_{D} \ket{j}_{R}.
\label{bound}
\end{equation}
Thus an upper limit on the confidence with which it is possible to identify state $\ket{\psi_{j}}_{L}$ is given by the overlap of the corresponding state $\ket{j}_{R}$ of the right system with $\mathcal{H}_{RD}$.  Using Eq.\ \ref{PD} we now obtain
\begin{equation}
\begin{array}{ccl}
[P(\psi_{j}|\omega_{j})]_{max} &=& _{R}\bra{j} {\rm Tr}_L (\hat{\rho}_L^{-1/2} \ket{\Psi} \bra{\Psi} \hat{\rho}_L^{-1/2}) \ket{j}_R \\
&=& {\rm Tr}_L (\hat{\rho}_L^{-1/2} {}_R\braket{j}{\Psi} \braket{\Psi}{j}_R \hat{\rho}_L^{-1/2}) \\
&=& p_j {\rm Tr}_L (\hat{\rho}_{jL} \hat{\rho}_L^{-1})
\end{array}
\label{limit}
\end{equation}
where $\hat{\rho}_{jL} = \ket{\psi_{j}}_{LL}\bra{\psi_{j}}$.  This bound agrees with that given in \cite{maxconf}, and has been reached experimentally for a set of three elliptical polarisation states \cite{expt}.

To derive the optimal POM we note from Eq.\ \ref{rhoRj} that the limit is achieved if $\hat{\rho}_{R|j} \propto \hat{P}_D \ket{j}_{R \, R} \bra{j} \hat{P}_D$.  Using Eq.\ \ref{PD} we see that
\begin{equation}
\begin{array}{ccl}
P_D \ket{j}_R &=& {\rm Tr}_L \left(\hat{\rho}_L^{-1/2} \ket{\Psi} \bra{\Psi} \hat{\rho}_L^{-1/2} \right) \ket{j}_R \\
&=& {\rm Tr}_L \left(\hat{\rho}_L^{-1/2} \ket{\Psi} (p_j^{1/2} {}_L \bra{\psi_j} \hat{\rho}_L^{-1/2}) \right) \\
&=& p_j^{1/2} {}_L \bra{\psi_j} \hat{\rho}_L^{-1} \ket{\Psi}
\end{array}
\end{equation}
and thus
\begin{equation}
\begin{array}{ccl}
\hat{\rho}_{R|j} &\propto& p_j {}_L \bra{\psi_j} \hat{\rho}_L^{-1} \ket{\Psi} \bra{\Psi} \hat{\rho}_L^{-1} \ket{\psi_j}_L \\
&=& {\rm Tr}_L \left( \ket{\Psi} \bra{\Psi} \hat{\rho}_L^{-1} p_j \hat{\rho}_{jL} \hat{\rho}_L^{-1} \right).
\end{array}
\end{equation}
Suppose now that the measurement outcome $\omega_j$ is associated with the POM element $\hat{\Pi}_j$ acting on the left system.  Then
\begin{equation}
\hat{\rho}_{R|j} \propto {\rm Tr}_L (\ket{\Psi} \bra{\Psi} \hat{\Pi}_j),
\label{Pij}
\end{equation}
and thus
\begin{equation}
\hat{\Pi}_j \propto \hat{\rho}_L^{-1} p_j \hat{\rho}_{jL} \hat{\rho}_L^{-1}.
\label{Pijpure}
\end{equation}
This reproduces the result in \cite{maxconf}, but has arisen here as a direct consequence of the no-signalling condition.  It is a feature of the maximum confidence strategy that the optimal POM elements are only defined up to an arbitrary constant, as may be seen from Eq.\ \ref{Pijpure}.  Note that these POM elements may not form a complete measurement, and in general an inconclusive result is also needed \cite{maxconf}.

For mixed states $\{ \hat{\rho}_{iL} \}$ similar results can be obtained.  In this case we consider the entangled state
\begin{equation}
\ket{\Psi} = \sum_{i \in \cup_{j} \sigma(j)} \beta_{i}^{1/2} \ket{\beta_{i}}_{L} \ket{i}_{R},
\label{psimixed}
\end{equation}
where $\sigma(j)$ is an index set associated with $\hat{\rho}_{jL}$, ${}_{L}\braket{\beta_{i}}{\beta_{i}}_{L} = 1$, and $\sum_{i \in \sigma(j)} \beta_{i} = p_{j}$, such that $p_{j} \hat{\rho}_{jL} = \sum_{i \in \sigma(j)} \beta_{i} \ket{\beta_{i}}_{LL} \bra{\beta_{i}}$.  The maximum confidence and optimal POM element may be derived using \emph{any} valid decomposition of $p_{j} \hat{\rho}_{jL}$, and $\{ \ket{\beta_{i}}_{L} \}$ need not be the eigenkets of $\hat{\rho}_{jL}$.  We will not discuss the full derivation for the general case, as it follows the same arguments outlined above, but proceed instead to illustrate both the pure state case, and the extension of the arguments to include mixed states, by means of an example.

Consider therefore the example of one mixed state and one pure state in a two-dimensional space, given by:
\begin{equation}
\begin{array}{ccl}
\hat{\rho}_{0} &=& q \ket{0} \bra{0} + (1-q) \ket{1} \bra{1}, \\
\hat{\rho}_{1} &=& \ket{\psi} \bra{\psi} = \frac{1}{2} (\ket{0} + \ket{1})(\bra{0} + \bra{1}),
\end{array}
\end{equation}
with \emph{a priori} probabilities $p$ and $1-p$ respectively.  Following Eq.\ \ref{psimixed}, we construct the entangled state:
\begin{equation}
\begin{array}{ccl}
\ket{\Psi} &=& (pq)^{1/2} \ket{0}_{L} \ket{0}_{R} + (p(1-q))^{1/2} \ket{1}_{L} \ket{1}_{R} \\
&& + (1-p)^{1/2} \ket{\psi}_L \ket{2}_{R}.
\end{array}
\end{equation}
We first consider the projector $\hat{P}_D$.  It is clear that $\braket{\Psi}{\Psi^{\perp}}=0$ where
\begin{equation}
\begin{array}{ccl}
\ket{\Psi^{\perp}} &=& k ((1-p)^{1/2} ((1-q)^{1/2} \ket{0}_{R}  + q^{1/2} \ket{1}_{R}) \\
&& - (2pq(1-q))^{1/2} \ket{2}_{R}),
\end{array}
\end{equation}
and $k=(1-p+2pq(1-q))^{-1/2}$ is a normalisation factor chosen to ensure $\braket{\Psi^{\perp}}{\Psi^{\perp}} = 1$.  Thus
\begin{equation}
\hat{P}_{D} = \hat{I}_R - \ket{\Psi^{\perp}}\bra{\Psi^{\perp}}.
\end{equation}
The probability that the state of the left system was $\hat{\rho}_0$ is equivalent to the probability that the right system lies in the subspace with the projector $\ket{0}_{R \, R} \bra{0} + \ket{1}_{R \, R} \bra{1}$ after measurement.  Thus for state $\hat{\rho}_0$, we obtain
\begin{equation}
\begin{array}{ccl}
P(\hat{\rho}_0 | \omega_0) &=& {\rm Tr}_L \left( \hat{\rho}_{R|0} \big( \ket{0}_{R \, R} \bra{0} + \ket{1}_{R \, R} \bra{1} \big) \right) \\
&=& {\rm Tr}_L \left( \hat{P}_D \hat{\rho}_{R|0} \hat{P}_D \big( \ket{0}_{R \, R} \bra{0} + \ket{1}_{R \, R} \bra{1} \big) \right) \\
&=& {\rm Tr}_L \left( \hat{\rho}_{R|0} \big(\hat{P}_D \big(\ket{0}_{R \, R} \bra{0} + \ket{1}_{R \, R} \bra{1} \big) \hat{P}_D \big) \right) \\
&\leq& \gamma_{max} \left(\hat{P}_D \big( \ket{0}_{R \, R} \bra{0} + \ket{1}_{R \, R} \bra{1} \big) \hat{P}_D \right), \\
\end{array}
\label{mixedbound}
\end{equation}
where we have used $\hat{\rho}_{R|0} = \hat{P}_D \hat{\rho}_{R|0} \hat{P}_D$, as $\hat{\rho}_{R|0}$ lies in the subspace with the projector $\hat{P}_D$, and $\gamma_{max}(\hat{X})$ denotes the largest eigenvalue of $\hat{X}$.  Note that $\ket{\phi_0} = q^{1/2} \ket{0}_{R} - (1-q)^{1/2} \ket{1}_{R}$ is an eigenket with eigenvalue 1 of both $\hat{P}_{D}$ and $\ket{0}_{RR}\bra{0} + \ket{1}_{RR} \bra{1}$.  Thus this limit is 1, corresponding to unambiguous discrimination, as expected, and the limit is reached if $\hat{\rho}_{R|0} = \ket{\phi_0} \bra{\phi_0}$.  We can readily deduce the optimal POM element $\hat{\Pi}_0$ by noting that ${}_R \bra{2} \hat{\rho}_{R|0} \ket{2}_R = 0$, and therefore, according to Eq.\ \ref{Pij} we require
\begin{equation}
_R \bra{2} {\rm Tr}_L(\ket{\Psi} \bra{\Psi} \hat{\Pi}_0)  \ket{2}_R = (1-p)_L\bra{\psi} \hat{\Pi}_0 \ket{\psi}_L = 0.
\end{equation}
Thus $\hat{\Pi}_0$ is a projector onto the state orthogonal to $\ket{\psi}$, also as expected.

For result $\omega_1$, using Eq.\ \ref{bound}, we obtain
\begin{equation}
P(\hat{\rho}_{1}|\omega_{1}) \leq 1 - |\braket{\Psi^{\perp}}{2}_{R}|^{2} = \frac{1-p}{1-p+2pq(1-q)}.
\end{equation}
This limit is achieved if $\hat{\rho}_{R|1} \propto \hat{P}_D \ket{2}_{R \, R} \bra{2} \hat{P}_D$.  Thus, from Eq.\ \ref{Pij} we require
\begin{equation}
{\rm Tr}_L(\ket{\Psi} \bra{\Psi} \hat{\Pi}_1) = c_1 \hat{P}_D \ket{2}_{R \, R} \bra{2} \hat{P}_D
\end{equation}
where we have introduced the constant of proportionality $c_1$.  We now note that we can easily write down the matrix elements of $\hat{\Pi}_1$ in the basis $\{ \ket{0}_L, \ket{1}_L \}$ as follows.  Clearly
\begin{eqnarray}
{}_L \bra{0} \hat{\Pi}_1 \ket{0}_L &=& \frac{{}_R \bra{0} {\rm Tr}_L(\ket{\Psi} \bra{\Psi} \hat{\Pi}_1) \ket{0}_R}{pq} \nonumber \\
&=& c_1 \frac{{}_R \bra{0} \hat{P}_D \ket{2}_{R \, R} \bra{2} \hat{P}_D \ket{0}_R}{pq} \\
&=& c_1 \frac{|_R\braket{0}{\Psi^{\perp}}|^2 |_R \braket{2}{\Psi^{\perp}}|^2}{pq} \nonumber \\
&=& c_1 k^4 2 (1-p) (1-q)^2. \nonumber
\end{eqnarray}
Similarly we can show
\begin{eqnarray}
{}_L \bra{0} \hat{\Pi}_1 \ket{1}_L = {}_L \bra{1} \hat{\Pi}_1 \ket{0}_L &=& c_1 k^4 2 (1-p) q (1-q) \quad \\
{}_L \bra{1} \hat{\Pi}_1 \ket{1}_L &=& c_1 k^4 2 (1-p) q^2.
\end{eqnarray}
As $\{ \ket{0}_L, \ket{1}_L \}$ forms an orthonormal basis for the left system, the above completely specifies the probability operator $\hat{\Pi}_1$, which is given by
\begin{equation}
\begin{array}{ccl}
\hat{\Pi}_1 &=& 2 c_1 k^4 (1-p) \left( (1-q) \ket{0}_L + q \ket{1}_L \right) \\
&& \left( (1-q) {}_L \bra{0} + q {}_L \bra{1} \right).
\end{array}
\end{equation}
Note that we have derived the bounds and optimal POM elements without needing to diagonalise $\hat{\rho}_L$, which in this case would be tedious (but straight-forward).  For higher dimensional systems, as it becomes more difficult to diagonalise $\hat{\rho}_L$, this approach may therefore be useful.

The above arguments allow us to interpret the bound solely in terms of the geometry of the right system, from which some results follow naturally.  The restriction imposed by no signalling ensures that the right system is confined to a $D$-dimensional subspace $\mathcal{H}_{AD}$, and clearly the bounds in Eqs. \ref{bound} and \ref{mixedbound} depend only on the overlap between this subspace and the basis states $\ket{i}_R$ of the right system.  We first note that for linearly independent pure states, $D=N$, and the right system occupies the entire space available.  The limit is therefore unity, corresponding to unambiguous or error-free discrimination \cite{unamb}.

Consider now the effect on these bounds of a transformation of the left system $\{ \hat{\rho}_{iL}, p_{i} \} \rightarrow \{ \hat{\rho}_{iL}^{\prime}, p_{i}^{\prime} \}$.  Suppose that we have a transformation conditioned on the outcome of some measurement, and that for a given measurement result the associated Kraus operator \cite{kraus} is $\hat{A}_{L}$.  It then follows that
\begin{eqnarray}
\hat{\rho}_{iL}^{\prime} &=& \frac{\hat{A}_{L} \hat{\rho}_{iL} \hat{A}_{L}^{\dagger}}{{\rm Tr}(\hat{\rho}_{iL} \hat{A}_{L}^{\dagger} \hat{A}_{L})}, \\
p_{i}^{\prime} &=& p_{i} {\rm Tr}(\hat{\rho}_{iL} \hat{A}_{L}^{\dagger} \hat{A}_{L}),
\end{eqnarray}
where $\hat{A}_L$ is chosen such that ${\rm Tr}(\hat{\rho}_{L} \hat{A}_{L}^{\dagger} \hat{A}_{L}) = 1$.  If $\hat{A}_{L}$ has rank $M \leq D$, the transformed set occupies an $M$-dimensional subspace.  By considering again the Schmidt decomposition of the transformed entangled state $\hat{A}_{L} \ket{\Psi}$, we can see that the right system must also now occupy an $M$-dimensional subspace $\mathcal{H}_{RM}$.  No signalling requires that $\mathcal{H}_{RM} \subseteq \mathcal{H}_{RD}$, with equality if and only if $M = D$.  But the maximum confidence depends only on the overlap of the projector $\sum_{i \in \sigma(j)} \ket{i}_{RR} \bra{i}$ with the subspace $\mathcal{H}_{RD}$ (respectively $\mathcal{H}_{RM}$ for the transformed system).  Thus it is clear that $[P(\hat{\rho}_{jL}^{\prime} | \omega_{j})]_{max} \leq [P(\hat{\rho}_{jL} | \omega_{j})]_{max}$, i.e. no physically allowed transformation, deterministic or probabilistic, of a given set of states $\{ \hat{\rho}_{i} \}$ with associated probabilities $\{p_{i} \}$ can result in a set which is more distinguishable by this measurement strategy.  In particular, for a rank $D$ operator (i.e. one describing an invertible transformation), this figure of merit is identical for the transformed and untransformed sets, as pointed out in \cite{maxconf}.

Finally, we note that there is an relationship between entanglement concentration and maximum confidence discrimination, analogous to that which exists for unambiguous discrimination \cite{che}.  Unambiguous discrimination may be thought of as occuring in two steps - transformation of the states into mutually orthogonal ones with some probability of success, followed by a von Neumann measurement to discriminate perfectly between the transformed states.  If the transformation fails, the outcome is interpreted as inconclusive.  It turns out that the optimal transformation (that minimising the probability of an inconclusive result), is also the one which transforms an appropriately chosen non-maximally entangled state to a maximally entangled one, with the optimal probability of success \cite{che}.  The maximum confidence measurement may also be thought of as a two-step process, in a way completely analogous to that for unambiguous discrimination \cite{maxconf}.  In the first step, the states are transformed according to $\hat{\rho}_i \rightarrow \frac{1}{D} \hat{\rho}^{-1/2} \hat{\rho}_i \hat{\rho}^{-1/2}$ with some probability of success $p_{succ}$.  If this fails, the outcome may be interpreted as inconclusive, and is associated with a probability operator $\hat{\Pi}_{fail} = \hat{I} - \frac{p_{succ}}{D} \hat{\rho}^{-1}$.  If it succeeds, the second step consists of a measurement corresponding to projectors onto the new states (for pure states).  The probabilty of success is limited by the condition $\hat{\Pi}_{fail} \geq 0$.  Thus
\begin{equation}
\hat{I} - \frac{p_{succ}}{D} \hat{\rho}^{-1} = \sum_{i=0}^{D-1} \left(1 - \frac{p_{succ}}{\lambda_i D} \right) \ket{\lambda_i} \bra{\lambda_i} \geq 0,
\end{equation}
and we require $p_{succ} \leq \lambda_i D$, $\forall \, \lambda_i$.  The optimal probability of success for the transformation $\hat{\rho}_i \rightarrow \frac{1}{D} \hat{\rho}^{-1/2} \hat{\rho}_i \hat{\rho}^{-1/2}$ is therefore $\lambda_{min} D$.  Consider now the effect of this transformation on the entangled state $\ket{\Psi}$ from Eq.\ \ref{schmidt}.
\begin{equation}
\ket{\Psi} \rightarrow \frac{1}{\sqrt{D}} \hat{\rho}_L^{-1/2} \ket{\Psi} = \frac{1}{\sqrt{D}} \sum_{i=0}^{D-1} \ket{\lambda_i}_L \ket{\lambda_i}_R.
\end{equation}
This operation transforms the non-maximally entangled state $\ket{\Psi}$ into a maximally entangled one.  Note that the optimal probability of success, $p_{succ} = \lambda_{min} D$, reaches the bound given in \cite{lo01} for transforming a non-maximally entangled state to a maximally entangled one by local operations and classical communication.

The no signalling condition is fundamental in reconciling quantum mechanics and special relativity, and there are now several examples of this condition being employed as a physical law to limit quantum mechanical operations.  In this paper we have used no-signalling as a principle to limit quantum state discrimination, and have shown that these considerations lead naturally to the maximum confidence strategy.  We have shown explicitly that for pure states the bound and optimal measurement may be derived from no-signalling.  We have also discussed how this may be extended to mixed states, and illustrated the argument by means of an example.  The no-signalling argument provides a direct way to find the bound and optimal measurement, without the need to diagonalise the density operator, as would otherwise be required.  No-signalling thus provides an alternative approach, both enhancing our understanding of this strategy, and serving as a calculational tool.  Some results arise naturally from this approach, notably that the bound cannot increase under any transformation, deterministic or probabilistic, of the states.  We also demonstrated the link between maximum confidence measurements and entanglement concentration.  The manipulation of entanglement is of considerable interest at the present time due to its importance in quantum information processing \cite{nie}.

\begin{acknowledgments}  We thank J. Jeffers and C. Gilson for helpful discussions.  S. C. acknowledges support by the Synergy fund of the Universities of Glasgow and Strathclyde.  E. A. acknowledges financial support from the Royal Society.  S. M. B. thanks the Royal Society and the Wolfson Foundation for support.  \end{acknowledgments}


\begin{thebibliography}{99}
\bibitem{asp} A. Aspect, Nature, \textbf{398}, 189 (1999)
\bibitem{ghi} C. G. Ghirardi, A. Rimini and T. Weber, Lett. Nuovo Cimento Soc. Ital. Fis. \textbf{27}, 293 (1980); P.J Bussey, Phys. Lett. A \textbf{90}, 9 (1982); T. F. Jordan, Phys. Lett. A \textbf{94}, 264 (1983)
\bibitem{clone} N. Gisin, Phys. Lett. A, \textbf{242}, 1 (1998); S. Ghosh, G. Kar, and A. Roy, Phys. Lett. A \textbf{261}, 17 (1999)
\bibitem{amb} S. M. Barnett and E. Andersson, Phys. Rev. A, \textbf{65}, 044307 (2002); D. Qui, Phys. Lett. A, \textbf{303}, 140 (2002); Y. Feng, S. Zhang, R. Duan and M. Ying, Phys. Rev. A, \textbf{66}, 062313 (2002)
\bibitem{pr} S. Popescu and D. Rohrlich, Found. Phys. \textbf{24}, 379 (1994)
\bibitem{mas} Ll. Masanes, A. Acin, and N. Gisin, Phys. Rev. A \textbf{73}, 012112 (2006)
\bibitem{barr} J. Barrett, L. Hardy and A. Kent, Phys. Rev. Lett. \textbf{95}, 010503 (2005)
\bibitem{hwang} W.-Y. Hwang, Phys. Rev. A, \textbf{71}, 062315 (2005)
\bibitem{maxconf} S. Croke, E. Andersson, S. M. Barnett, C. R. Gilson and J. Jeffers, Phys. Rev. Lett. \textbf{96}, 070401 (2006)
\bibitem{hel} C. W. Helstrom, \emph{Quantum Detection and Estimation Theory}, Academic, New York (1976)
\bibitem{nie} M. A. Nielsen and I. L. Chuang, \emph{Quantum Computation and Quantum Information}, Cambridge University Press, Cambridge (2000)
\bibitem{expt} P. J. Mosley, S. Croke, I. A. Walmsley and S. M. Barnett, Phys. Rev. Lett. \textbf{97}, 193601 (2006); S. Croke, P. J. Mosley, S. M. Barnett and I. A. Walmsley, Eur. Phys. J. D \textbf{41}, 589 (2007)
\bibitem{unamb} I. D. Ivanovic, Phys. Lett. A \textbf{123}, 257 (1987); D. Dieks, Phys. Lett. A, \textbf{126}, 303 (1988); A. Peres, Phys. Lett. A, \textbf{128}, 19 (1988)
\bibitem{kraus} K. Kraus, \textit{States, Effects and Observations, }Lecture
Notes in Physics \textbf{190}, Springer Verlag, Heidelberg (1983)
\bibitem{che} A. Chefles and S. M. Barnett, Phys. Lett. A \textbf{236}, 177 (1997); A. Chefles, Phys. Lett. A, \textbf{239}, 339 (1998)
\bibitem{lo01} H.-K. Lo and S. Popescu, Phys. Rev. A \textbf{63}, 022301 (2001)
\end{thebibliography}
\end{document}